\documentclass[conference]{IEEEtran}
\usepackage{cite}
\usepackage{amsmath,amssymb,amsfonts}
\usepackage{algorithm}
\usepackage{algpseudocode}
\usepackage{graphicx}
\usepackage{textcomp}
\usepackage{xcolor}
\usepackage{circledsteps}

\def\BibTeX{{\rm B\kern-.05em{\sc i\kern-.025em b}\kern-.08em
    T\kern-.1667em\lower.7ex\hbox{E}\kern-.125emX}}
    
\begin{document}

        \title{Locality, Latency and Spatial-Aware Data Placement Strategies at the Edge\\
        }

        \author{
            \IEEEauthorblockN{Nikhil Sreekumar, Abhishek Chandra, Job B. Weissman}
            \IEEEauthorblockA{
                \textit{Department of Computer Science and Engineering}\\
                \textit{University of Minnesota}\\
                Minneapolis, MN \\
                \{sreek012, chandra\}@umn.edu, \{jon\}@cs.umn.edu
            }
        }

        \maketitle

        \begin{abstract}
            The vast data deluge at the network's edge is raising multiple challenges for the edge computing community. One of them is identifying edge storage servers where data from edge devices/sensors have to be stored to ensure low latency access services to emerging edge applications. Existing data placement algorithms mainly focus on locality, latency, and zoning to select edge storage servers under multiple environmental constraints. This paper uses a data placement framework to compare distance-based, latency-based, and spatial-awareness-based data placement strategies, which all share a decision-making system with similar constraints. Based on simulation experiments, we observed that the spatial-awareness-based strategy could provide a quality of service on par with the latency-based and better than the distance-based strategy.
        \end{abstract}


        \section{Introduction}
    \label{sec:intro}
    The International Data Corporation (IDC) expects 55.9 billion connected devices by 2025, generating 79.4ZB data \cite{bib:idc}. This data deluge is increasing the data gravity at the network's edge resulting in the emergence of edge applications\cite{bib:seminal}. Deploying applications near the source of data generation ensures low data access latency and low service latency with improved quality of service. For example, in the case of AR/VR applications, the MTP(Motion to Photon) latency should be less than 20 ms for immersive experience \cite{bib:arvr}. Most of the data generated can be stored at the edge, utilized by edge applications, and then either sent to the cloud for persistent storage or discarded. This temporary buffering strategy can decrease the traffic congestion on the wide-area backhaul link while giving data owners more control to choose what data to be stored on the public cloud. However, there are multiple challenges one should address to ensure the quality of service to users when it comes to storing data at the edge \cite{bib:sharecare,bib:edgenativestore}. The heterogeneity of edge storage nodes, limited elasticity, node churn, user mobility, time-sensitive compute-value of data, and privacy opens multiple doors to edge storage research. Furthermore, with the introduction of compliance laws like GDPR \cite{bib:gdpr}, it is essential for data (for example, health and trading) to be stored persistently within a region.

    The storage nodes at the edge are heterogeneous in terms of storage capacity, network bandwidth, and request-handling power. The limited elasticity and node churn make it even more challenging. The placement of data under these environmental conditions to provide the best quality of service for application users by minimizing the average latency is not a trivial task. We will call data generation sources, Producers; data consuming applications, Consumers, and edge storage servers, Hosts in the rest of the paper. Multiple consumers can subscribe to the data generated by producers. Also, a single consumer can subscribe to multiple producer data. Depending on the consumer demand and the available storage capacity, new replica hosts may have to be created on the fly to continue service. Existing data placement systems focus on single replication \cite{bib:ifogstor} and multi-replication \cite{bib:ifogstorm, bib:ifogstors} of data using distance, latency and geo-aware \cite{bib:coin, bib:datapop} strategies. However, the host request handling capacity needs to be considered while making data placement decisions. Also, using location, latency, and spatial awareness separately may increase the end-to-end latency and decision-making time.
    
    \pgfkeys{/csteps/fill color=white}
    We propose the following enhancements that can significantly impact the quality of service: \Circled{1} The data placement decision should consider the ingress and egress load capacity per host node for replica selection (Producer/Consumer load constraint in Section \ref{subsec:prob}). \Circled{2} The density of producers and consumers in a region should be considered for selecting a replica (Centroid-based host selection in Section \ref{sec:heur}). \Circled{3} A combined use of geolocation, latency, and spatial awareness can prune the potential host search space, resulting in less decision-making time (Spatial-awareness-based strategy in Section \ref{sec:heur}).
    
    We propose a data placement framework that accommodates the above enhacements and explores the following questions:
    \begin{itemize}
        \item How will the distance-based, and latency-based selection of hosts for data placement affect end-to-end latency observed by consumers?
        \item Can we combine the features of distance-based and latency-based along with spatial awareness to create a better data placement policy?
        \item With varying application workloads, how will the average number of replicas generated per producer vary across all three strategies?
        \item With an increasing number of hosts, producers, and consumers, can the distance-based, latency-based, and spatial-awareness-based strategies scale?
    \end{itemize}

    The main contributions in this paper are
    \begin{itemize}
        \item Proposed a problem formulation for data placement that considers the storage capacity, producer/consumer load threshold, and on-demand replication of hosts to minimize the overall average end-to-end latency observed by consumers.
        \item Proposed a data placement framework that uses three strategies: distance-based, latency-based, and spatial-awareness-based under similar system constraints.
        \item All three strategies are compared using simulation experiments where we made the following observations: \Circled{1} The use of distance to identify a potential set of host nodes may only be sometimes beneficial as the selected nodes incur high latency. \Circled{2} Combining the properties of location-based and latency-based strategies can improve the latency observed by consumers. \Circled{3} Spatial awareness can be used to prune the potential host search space, reducing the decision-making time for replica creation which can scale with increasing consumer demand.
    \end{itemize}

    Section \ref{sec:sysmodel} gives a brief overview of the system model and data placement formulation. The proposed data placement framework that uses the three data placement strategies is described in Section \ref{sec:heur}. A comparison across the data placement strategies is presented in Section \ref{sec:eval} using simulation experiments. We conclude our work with a brief discussion of future work and the major findings in the paper in Section \ref{sec:conclude}.
        \section{System Model}
    \label{sec:sysmodel}
    \begin{figure}
        \centering
        \includegraphics[scale=0.2]{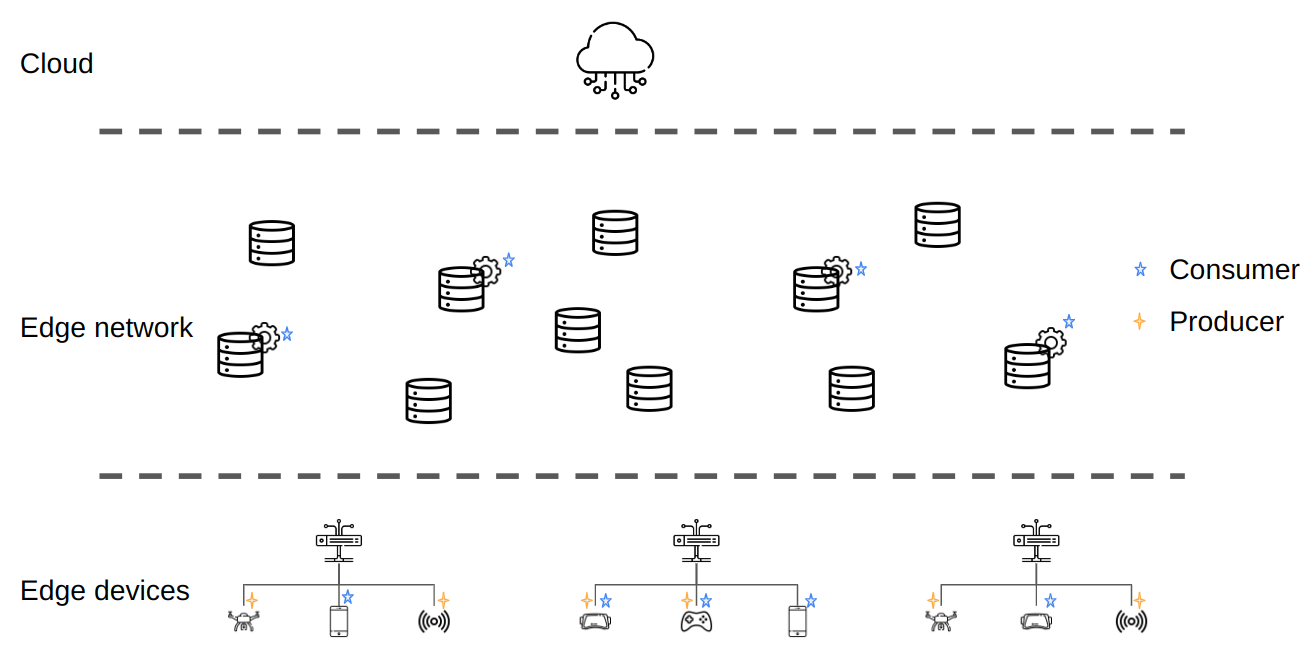}
        \caption{System architecture}
        \label{fig:sys_arch}
    \end{figure}

\begin{figure}
        \centering
        \includegraphics[scale=0.5]{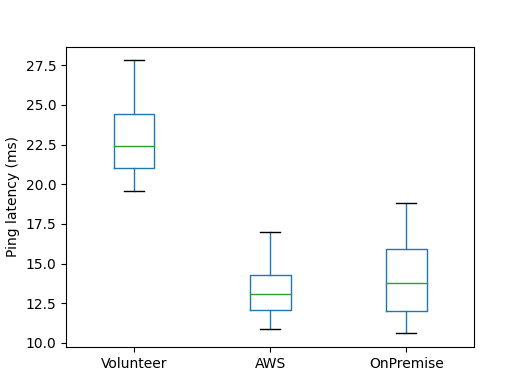}
        \caption{Ping latency test. AWS local server and on-premise servers have lower ping latency compared to volunteer nodes}
        \label{fig:ping_lat}
    \end{figure}

    The system architecture is shown in Figure \ref{fig:sys_arch}. There are a set of data generators (producers), a set of edge storage nodes (hosts), and a set of application tasks (consumers) in the system. Producers generate data and store/stream it to a host, where a consumer will collect the data. Gateways allow the connection of producers to the internet and vice versa. Multiple consumers can subscribe to a producer's data, and consumers can subscribe to multiple producer data. The hosts vary in storage capacity and ingress and egress request handling capacity. Based on a ping latency experiment (Figure. \ref{fig:ping_lat}) on AWS local servers, on-premise servers, and volunteers' servers, we could see that high-resource servers (AWS and on-premise servers) have low latency. We are focusing on a setting with latency directly proportional to resource capacity, as this is one plausible scenario. In future work, we plan to identify how the different strategies will perform in different network and resource settings. A producer can directly communicate with a host once it is identified as a suitable location for data storage. A host can communicate with any other host node within a region.

    The producer generates data and sends it to a suitable host for storage. If there are pending consumer requests at the host, the data is stored and immediately shared with the consumers. We consider long-running edge application services, for example, a car insurance company perusing accident videos for validating insurance claims or a city planning system that wants to rearrange the traffic over coming days based on hourly data. It can also be for applications like an augmented reality-based robotic surgery handled by multiple surgeons that require the same data to be streamed to all involved doctors to avoid life-threatening situations. This AR/medical data will have to be stored at the host node in case someone wants to check back on a procedure carried out sometime back. There can be two types of latency here: data retrieval latency(car insurance or city planning where data is already present at the host)  or end-to-end latency(augmented reality surgery where data is streaming). In this paper, we will focus on end-to-end latency scenarios.

    \subsection{Problem Formulation}
    \label{subsec:prob}
        Given sets of Producers ($P$), Hosts ($H$) and Consumers ($C$) with sizes $p$, $h$ and $c$ respectively. Each producer $p_{i}$ sends data of size $datasize_{p_{i}}$ to at most $r$ hosts to meet the demands of subscribed consumers. The value of $r$ may vary from producer to producer. However, here we take $r = replica_threshold$ for all producers. A consumer ($c_{k}$) can subscribe to more than one producer. For simplicity, we assumed the data transfer unit is $b$ bytes. Each host has a storage capacity of $cap_{h_{j}}$, a producer load threshold $pload_{h_{j}}$ (number of concurrent producer connections), and a consumer load threshold $cload_{h_{j}}$. Each consumer maintains a binary subscription list $csub$ of size ($1 \times p$), where, $csub_{ki} = 1$, if $c_{k}$ subscribes to $p_{i}$, otherwise $csub_{ki} = 0$. The binary matrix, $phc$ of size ($p \times h \times c$) is used to represent the paths from a producer to a consumer via a host. If $phc_{ijk} = 1$, there exists a path from producer ($p_{i}$) to consumer ($c_{k}$) via host ($h_{j}$), otherwise $phc_{ijk} = 0$.
        \begin{itemize}
            \item \textit{Load constraint}: A producer $p_{i}$ or a consumer $c_{j}$ can use a host $h_{j}$ only if the addition of a new connection is within the producer or consumer load threshold respectively. If $plt_{h_{j}}$ is the producer load threshold and $clt_{h_{j}}$ is the consumer load threshold for host $h_{j}$, then\\
                \begin{equation}
                    \sum_{i=1}^{p} (\sum_{k=1}^{c} phc_{ijk}>0) \le plt_{h_{j}} \tag{1} \label{eq:c1_1}
                \end{equation}
                \begin{equation}
                    \sum_{k=1}^{c} (\sum_{i=1}^{p} phc_{ijk}>0) \le clt_{h_{j}} \tag{2} \label{eq:c1_3}
                \end{equation}
            
            \item \textit{Storage constraint}: A host ($h_{j}$) can store data from a producer ($p_{i}$) only if $datasize_{p_{i}}$ is less than the available storage capacity $cap_{h_{j}}$ of the host.
                \begin{equation}
                    \sum_{i=1}^{p} datasize_{p_{i}} * (\sum_{k=1}^{c} phc_{ijk} > 0) \le cap_{h_j} \tag{3} \label{eq:c2}
                \end{equation}

            \item \textit{Single path constraint}: There exist a single path from a producer $p_{i}$ to a consumer $c_{k}$ via a host $h_{j}$, provided $c_{k}$ is subscribed to $p_{i}$, i.e., $csub_{ki} = 1$.
                \begin{equation}
                    \sum_{j=1}^{h} (phc_{ijk} * csub_{ki}) = 1 \tag{4} \label{eq:c3}
                \end{equation}

            \item \textit{On demand constraint}: The number of replicas alloted to a producer ($p_{i}$) can be greater than or equal to $1$. Once there are no more resources to share, any incoming replica request is declined.
                \begin{equation}
                    \sum_{j = 1}^{h} (\sum_{k = 1}^{c} phc_{ijk} > 0) \ge 1 \tag{5} \label{eq:c4}
                \end{equation}
        \end{itemize}

        Given the above constraints, we need to ensure that the selected path ($phc_{ijk}$) for data transfer from a producer ($p_{i}$) to a consumer ($c_{k}$) via a host ($h_{j}$) has low latency. If the latency of transferring $b$ units of data between $p_{i}$ and $h_{j}$ is $lat_{ij}$ and that between $h_{j}$ and $c_{k}$ is $lat_{jk}$, then the total latency is given by
            \begin{equation}
                lat\_phc_{ijk} = (lat_{ij} + lat_{jk}) * \frac{datasize_{p_{i}}}{b} * phc_{ijk} \tag{6} \label{eq:c5}
            \end{equation}

        \textbf{Objective}\\
            Our data placement algorithm aims to find the hosts where data from the producers can be stored for consumption by consumers while providing minimum average end-to-end latency. It means filling the binary matrix $phc$ by minimizing (\ref{eq:c5}) for all producers, hosts, and consumers.
            \begin{equation} \tag{7} \label{eq:c6}
                \begin{split}
                    Minimize \quad \quad            & \frac{\sum_{i=1}^{p} \sum_{j=1}^{h} \sum_{k=1}^{c} lat\_phc_{ijk}}{\sum_{i=1}^{p} \sum_{j=1}^{h} \sum_{k=1}^{c} phc_{ijk}}\\
                    subject \,\, to \quad \quad     & \sum_{i=1}^{p} (\sum_{k=1}^{c} phc_{ijk}>0) \le plt_{h_{j}}\\
                                                    & \sum_{k=1}^{c} (\sum_{i=1}^{p} phc_{ijk}>0) \le clt_{h_{j}}\\
                                                    & \sum_{i=1}^{p} datasize_{p_{i}} * (\sum_{k=1}^{c} phc_{ijk} > 0) \le cap_{h_j}\\
                                                    & \sum_{j=1}^{h} (phc_{ijk} * csub_{ki}) = 1\\
                                                    & \sum_{j = 1}^{h} (\sum_{k = 1}^{c} phc_{ijk} > 0) \ge 1
                \end{split}
            \end{equation}
    
    \subsection{Optimization Solution}
        The objective (\ref{eq:c6}) can be represented as a Mixed Integer Integer Programming problem. However, the formulation is similar to \cite{bib:ifogstor}, which is an NP-Hard problem. Therefore, as the number of actors in the system increases, the execution time will also increase, which is unsuitable for latency-sensitive applications. Hence, in the next section, we propose a data placement framework that uses three strategies to scale the solution.
        
        \section{Data Placement Framework}
\label{sec:heur}
\begin{figure}
            \centering
            \includegraphics[scale=0.2]{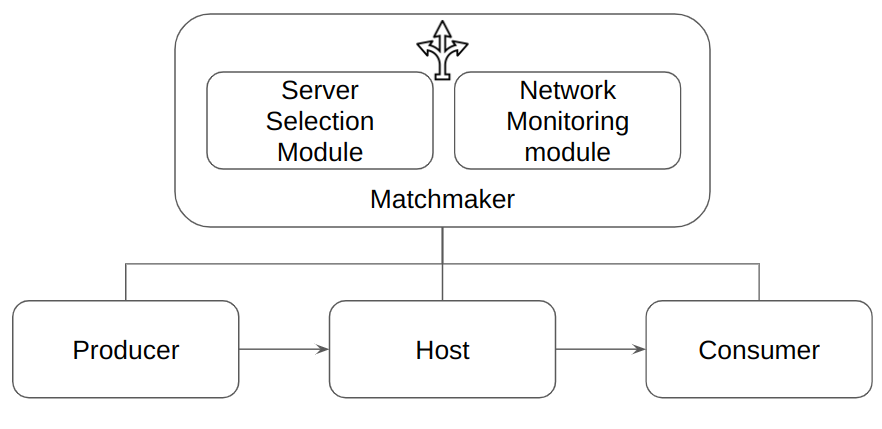}
            \caption{Matchmaker: The decision-making component}
            \label{fig:matchmaker}
        \end{figure}

    To scale the optimization problem with an increasing number of producers, hosts, and consumers, we propose a data placement framework that considers all the constraints mentioned in section \ref{subsec:prob}. The framework mainly consists of a central decision-making system called the Matchmaker, deployed on a dedicated, stable edge server. Upon entering the system, producers, hosts, and consumers will register with the Matchmaker. Producers share an estimated data size that will be sent to the selected host with the Matchmaker. Hosts share information on the producer load threshold, consumer load threshold, and total storage capacity. Finally, the consumer will share the subscription list. All three of them will also share their geo-location with the Matchmaker. The Matchmaker consists of a server selection module and a network monitoring module (Figure \ref{fig:matchmaker}). The server selection module selects the appropriate edge storage server according to distance-based, latency-based, or spatial-awareness-based data placement strategies. The network module monitors the network links across producers, hosts, and consumers periodically, the information from which is then used by the server selection module.
    
    \textbf{Initial Load conditioning}: Initially, when producers join the framework, the server selection module allocates no more than half the producer load per host node. This restriction is imposed to balance the load across the host nodes rather than focusing on a few ones. Suppose a host node with high capacity in terms of storage and producer load is selected for most producers. In that case, there is a chance that consumers of a few producers will take over the entire consumer load available in the host. This takeover can lead to storage space wastage and new replica creation overhead. The restriction enforced by the server selection module mitigates the takeover to some extent.

    \textbf{Centroid based host selection}: Once the initial load conditioning is complete, all producers will have a host allotted. As the demand increases, we will have to select a new host node for data replication. For the end-to-end latency scenario, producer and consumer information is essential to decide where data should be placed. Therefore, we consider the centroid of geo-locations of all existing consumers and the producer to select potential host nodes. This way, depending on the density of consumers, the host node selection will move towards locations with a high number of consumers. At the same time, we also take into account the producer. Once the centroid is identified, a data placement strategy is run on the potential host nodes to find a replica. The latency-based strategy does not require centroid-based selection as it always picks up the low latency, high resource-based host nodes for replication. Therefore, it is near-optimal among the three strategies.

    \subsection{Data Placement Strategies}
        \begin{enumerate}
            \item \textbf{Distance-based selection}: Host nodes that are at close physical proximity to the centroid is chosen to store the data from a producer.
            \item \textbf{Latency-based selection}: Host nodes at close network proximity to the producer/consumer is chosen. Centroid-based selection does not apply to latency-based strategy as it mostly picks the best host node with low latency and high capacity.
            \item \textbf{Spatial heuristic-based selection}: In a dense edge environment, using distance-based and latency-based selection may incur higher decision-making latency. The overhead can be decreased if we can somehow prune the search space. In this strategy, we prune the host search space using a spatial data structure, R-Tree \cite{bib:rtree}. The host nodes within the centroid's vicinity are first selected using the spatial data structure. Then, the best host with low latency to the producer and consumer is selected for data storage.
        \end{enumerate}

    In all the above strategies, we consider the load and storage capacity of the edge server before making the selection.

    \begin{figure}[H]
        \centering
        \includegraphics[scale=0.2]{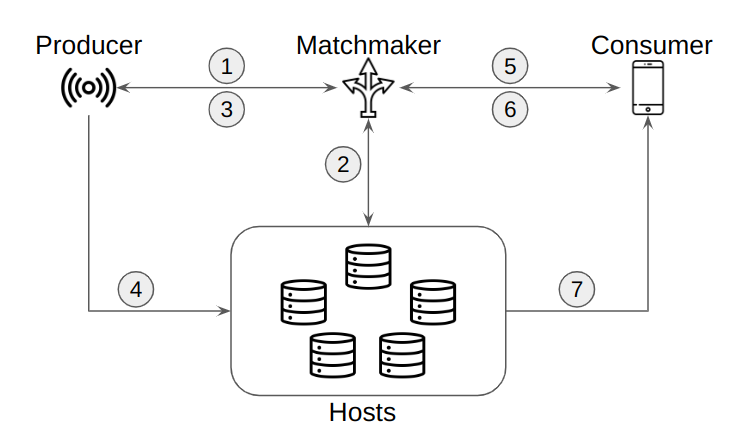}
        \caption{System workflow}
        \label{fig:workflow}
    \end{figure}

    \pgfkeys{/csteps/fill color=white}
    Consider the workflow in Figure. \ref{fig:workflow}. \Circled{1} The producers register with the Matchmaker by providing their location, id, and estimated storage size. The hosts will also send their location, storage capacity, and producer/consumer load threshold to the matchmaker \Circled{2}. The Matchmaker identifies the best host nodes and relays the information back to the producer \Circled{3}. More than one producer can store data on a host node. The producer can now directly contact the host to store the data \Circled{4}. The consumer registers with the Matchmaker by providing their location, application id, and subscription list \Circled{5}. Based on the subscription list, the Matchmaker identifies the best host nodes and shares the information with the consumer \Circled{6}. The consumer directly connects with the hosts for data transfer \Circled{7}.

        \begin{algorithm}
            \caption{Host Selection}
            \label{algo:hostselect}
            \begin{algorithmic}[1]
                \Procedure{SelectHost}{prod\_id, is\_replica, cons\_id}
                    \State host $\gets$ \{\}
                    \State loc $\gets$ \{\} \Comment{dummy location}
                    \If{is\_replica} 
                        \State centroid $\gets$ GetCentroid(prod\_id, cons\_id) \Comment{System model dependent}
                    \Else
                        \State loc $\gets$ GetLoc(prod\_id)
                    \EndIf
                    \State host $\gets$ DPStrategy(loc, prod\_id, cons\_id)
                    \State InformProducerofHost(prod\_id, host)
                    \If{is\_replica}
                        \State InformConsumerofHost(cons\_id, prod\_id, host)
                    \EndIf
                \EndProcedure
            \end{algorithmic}
        \end{algorithm}

        \pgfkeys{/csteps/fill color=white}
        The server selection module in the Matchmaker runs the data placement strategies in \textit{SelectHost} (algorithm \ref{algo:hostselect}) in two cases. \Circled[inner color=black]{a} Producer enters the system for the first time and \Circled[inner color=black]{b} Dynamic replication caused by overload on a host node.

        \pgfkeys{/csteps/fill color=white}
        In algorithm \ref{algo:hostselect}, depending on whether the procedure call is for a replica or initial host for the producer, the location of interest is calculated \Circled{line 4-8}. Based on the location, either distance-based or spatial-awareness-based strategy is called. For latency-based strategy, location is not used \Circled{line 9}. Once a host is selected, the producer and consumer are notified so that they can send and receive data respectively \Circled{line 10-13}. 

        \begin{algorithm}[t]
            \caption{Distance/Latency based Strategy}
            \label{algo:dlstrategy}
            \begin{algorithmic}[1]
                \Procedure{Distance/LatencyStrategy}{loc, prod\_id, cons\_id}
                    \State potential\_hosts $\gets$ SortHostByLocationOrLatency(loc, prod\_id, cons\_id)
                    \State host $\gets$ SelectBestViableHost(prod\_id, potential\_hosts)
                    \State return host
                \EndProcedure
            \end{algorithmic}
        \end{algorithm}

        For distance-based and latency-based strategies (algorithm \ref{algo:dlstrategy}), the Matchmaker first orders host nodes based on their sum of distance or latency from producer and consumer\Circled{line 2}. The best node is selected based on available storage capacity, and load \Circled{line 3}. In cases where all the resources are exhausted, the consumer replica requests are currently denied as immediate data sharing is not possible, given that applications like live streaming deem it necessary. The time complexity for distance/latency-based host selection is $O(nlogn)$, where $n$ is the number of host nodes under consideration. The network module provides latency information observed across the nodes.

\begin{figure*}
            \includegraphics[scale=0.2]{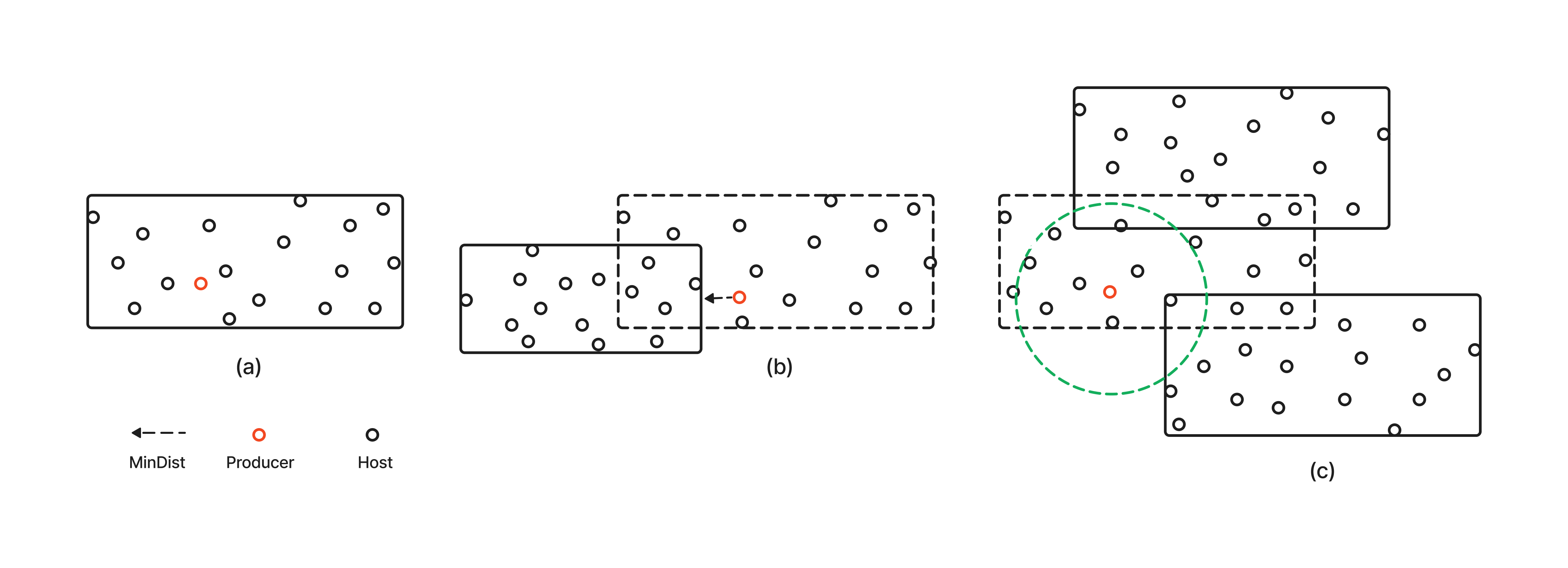}
            \caption{Host search (a) Search in MBR of centroid/producer (b) Search in all nearby MinDist MBR  (c) Search in unvisited MBRs expanding search concentrically from centroid/producer}
            \label{fig:hostsearch}
        \end{figure*}

        \pgfkeys{/csteps/fill color=white}
        For spatial-awareness-based strategy (algorithm \ref{algo:spatial}), the Matchmaker first checks if the loc is present inside one of the Minimum Bounding Rectangle (MBR) of the R-Tree. If present, the best host is selected and returned \Circled{lines 4-7}. Otherwise, the search extends to all the nearest MBRs identified using different MinDist \cite{bib:mindist} across the branches of the R-Tree. If present, the best host is selected and returned \Circled{lines 9-11}. If none of the mentioned searches find a host, the search incrementally starts from the $loc$ as concentric zones externally till the far host node is reached. If present, the best host is selected and returned \Circled{lines 9-11}. There is a chance that all the replicas are overloaded during this search. If so, the consumer requests are declined. The time complexity for the spatial-awareness-based strategy is $O(log_M (n))$, where $M$ is the maximum number of children per node in R-Tree and $n$ is the number of host nodes under consideration. A depiction of MinDist and Concentric Search is shown in Figure \ref{fig:hostsearch}.

        \begin{algorithm}
            \caption{Spatial Strategy}
            \label{algo:spatial}
            \begin{algorithmic}[1]
                \Procedure{SpatialStrategy}{loc, prod\_id, cons\_id}
                    \State host $\gets$ \{\}

                    \State
                    \State nodes $\gets$ SD.GetMBRNodes(loc) \Comment{SD is the RTree structure}
                    \If{$\neg$nodes.empty()}
                        \State host $\gets$ SelectHostFromCandidates(nodes, prod\_id, cons\_id)
                    \EndIf

                    \State
                    \If{host.empty()}
                        \State nodes $\gets$ SD.FindAllNearestMBRNodesByMinDist(loc)
                        \If{$\neg$nodes.empty()}
                            \State host $\gets$ SelectHostFromCandidates(nodes, prod\_id, cons\_id)
                        \EndIf
                    \EndIf
                    
                    \State
                    \If{host.empty()}
                        \State nodes $\gets$ \{\}
                        \While{nodes.empty() \& boundReached}
                            \State nodes $\gets$ SD.ConcentricSearch(id, loc, sz)
                        \EndWhile
                        \State host $\gets$ SelectHostFromCandidates(nodes, prod\_id, cons\_id)
                    \EndIf
                    
                    \State return host
                \EndProcedure
                \State
                \Procedure {SelectHostFromCandidates}{nodes, prod\_id, cons\_id}
                    \State potential\_hosts $\gets$ SortByLatency(loc, prod\_id, cons\_id)
                    \State host $\gets$ SelectBestViableHost(potential\_host, prod\_id)
                    \State return host
                \EndProcedure
            \end{algorithmic}
        \end{algorithm}

        \begin{algorithm}
            \caption{Host Selection for Consumer}
            \label{algo:chostselect}
            \begin{algorithmic}[1]
                \Procedure{ConsumerHostSelect}{cons\_id}
                    \State prods $\gets$ GetSubscribedList(cons\_id)
                    \For{prod\_id $\in$ prods}
                        \State nodes $\gets$ GetAssignedHosts(prod\_id)
                        \State nodes $\gets$ SortByDistanceOrLatency(nodes, cons\_id)
                        \State host $\gets$ SelectBestViableHostByLoad(nodes)
                        \If{host.empty()}
                            \State host $\gets$ SelectHost(prod\_id, true, cons\_id)
                        \EndIf
                    \EndFor
                    \State return host
                \EndProcedure
            \end{algorithmic}
        \end{algorithm}

    The Matchmaker will assign a host node for each consumer subscription by calling $ConsumerHostSelect$ (algorithm \ref{algo:chostselect}). Initially, the Matchmaker checks if any existing hosts allocated to a subscribed producer are available. The information of the available host with the shortest distance or latency is sent to the consumer \Circled{lines 4-6}, otherwise a call to the $SelectHost$ (algorithm \ref{algo:chostselect}) is made to create a new replica for the producer data \Circled{line 8}.
        \section{Evaluation}
    \label{sec:eval}

    \subsection{Experimental Setup}
        The simulation experiments are run on a Linux machine with 64GB RAM and 24 cores for the simulation. Based on the ping latency information in section \ref{sec:sysmodel},  we select 5ms-10ms, 10ms-15ms, and 15ms-20ms ranges for high-capacity storage, medium-capacity storage, and low-capacity storage host nodes. The location associated with producers, hosts, and consumers was taken from the Social IoT real-time dataset \cite{bib:socialiot}. The storage capacity of host nodes is in the range of 32GB-1TB (proportional to latency). Each host can support a load (producer + consumer) in the range of 40-80. The high-capacity host nodes have a higher load, followed by medium-capacity and low-capacity host nodes. The producer load is one-third of the total load, and the remaining is for the consumer load in the host. Producers can generate data in the range of 1GB-32GB. The consumer arrival follows Poisson distribution with a mean inter-arrival time of 5ms. Each producer will send chunks of size 1024 bytes to hosts until it reaches the data size to be generated. The RTree parameters $M$ (maximum number of children within a node) and $m$ (minimum number of children within a node) are set to 40 and 20, respectively.
    
    \subsection{Simulation experiments}

        \subsubsection{End-to-end latency}
\begin{figure}[t]
                \centering
                \includegraphics[scale=0.4]{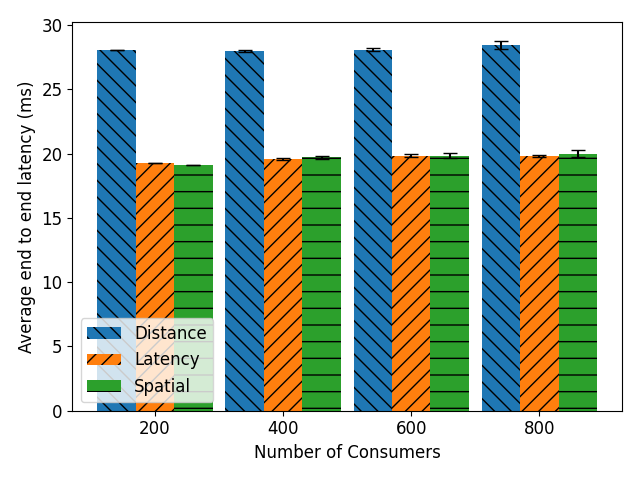}
                \caption{Average end-to-end latency. The number of hosts and producers is set to 50 and 100, respectively. Distance-based looks only at location information, leading to the selection of nodes with high latency. In contrast, latency-based and spatial-based consider latency resulting in low end-to-end latency.}
                \label{fig:str_e2e_lat}
            \end{figure}

            End-to-end latency provides a measure of the quality of service. In this experiment, we simulate 50 hosts, 100 producers, and a varying number of consumers (200-800). It can be seen in Figure. \ref{fig:str_e2e_lat}, the average end-to-end latency remains almost the same across all the difference (host, producer, consumer) configurations. This is because the number of times dynamic replication occurs is much less than the number of chunks transferred. A detailed look at the replica overhead is shown in section \ref{ssec:avg_ro}. The distance-based strategy takes more time as it does not consider the selected host's latency. There is also the chance that the host node has less resource capacity, leading to more replications. As for latency-based and spatial-based strategies, they can find the best host node within the search space in the given environmental setting. Therefore, the expectation would be to have less latency for spatial-based. However, the low number of hosts causes less contribution of host selection time at the Matchmaker to the average end-to-end latency. We will look at a scenario where spatial-based will outperform latency-based in section \ref{ssec:dec_time}.

        \subsubsection{Average replica count per producer}
\begin{figure}
                \centering
                \includegraphics[scale=0.4]{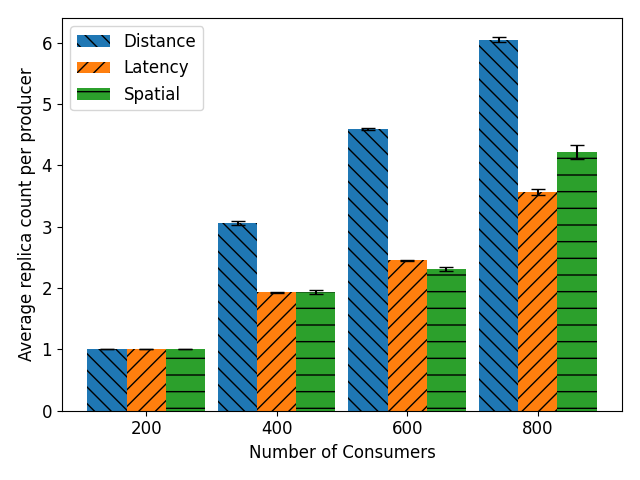}
                \caption{Average replica count per producer. The number of hosts and producers is set to 50 and 100 respectively. In the given environmental setting, distance-based may end up selecting hosts with less resource capacity. This selection leads to the creation of more replicas.}
                \label{fig:str_rep_count}
            \end{figure}

            For the same simulation scenario discussed above, the average count of replicas per producer is shown in Figure \ref{fig:str_rep_count}. It can be seen that latency-based and spatial-based outperform distance-based in all the configurations. Distance-based can choose host nodes that have less storage and less load threshold. This results in the creation of new replicas more often. Spatial-based shows a similar replica count to latency-based. However, there is a chance that spatial selects host nodes with comparatively fewer resources. This selection leads to more replicas, as shown in the configuration (50,100,800).
            
        \subsubsection{Average replication overhead per consumer}
            \label{ssec:avg_ro}
\begin{figure}
                \centering
                \includegraphics[scale=0.4]{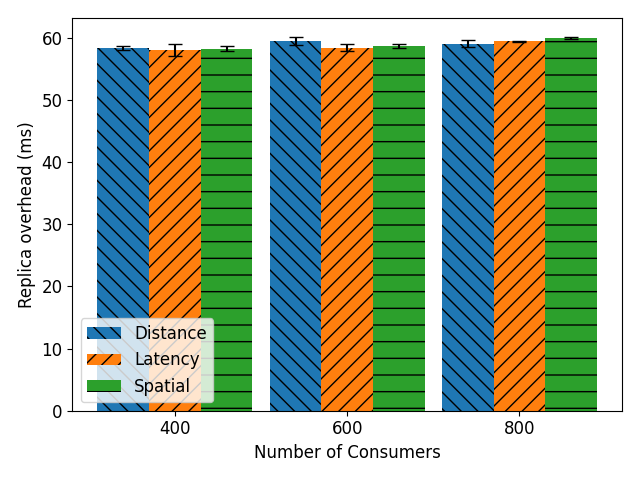}
                \caption{Average replication overhead per consumer. The number of hosts and producers is set to 50 and 100, respectively. Given the low host count, and the high number of chunks transferred, the average overhead across all the strategies is observed to be the same.}
                \label{fig:str_avg_ro}
            \end{figure}

\begin{figure}
                \centering
                \includegraphics[scale=0.4]{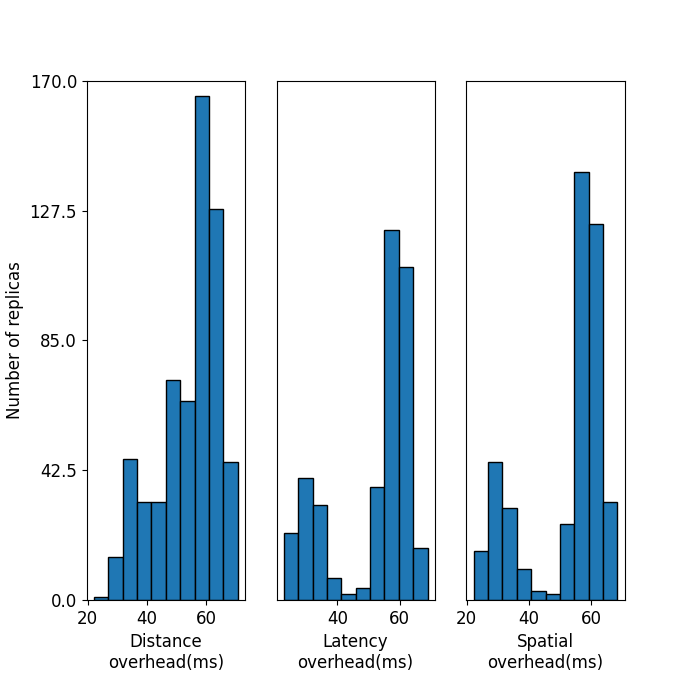}
                \caption{Replica overhead distribution. The replica overhead is distributed across all bins in distance-based, leading to similar overhead as latency-based and spatial-based.}
                \label{fig:str_ro_hist}
            \end{figure}

            The replication overhead is calculated as the elapsed time from when a consumer sends a request to the Matchmaker for a host corresponding to a producer till it receives the first chunk of data. When a replica is identified, the Matchmaker will request the producer to start a new connection with the replica to start sharing data. Once the host receives the data from the producer, it stores and immediately sends it to the consumer. In all three strategies, even though replication overhead is expected to vary, it is observed that the average overhead is almost the same (Figure \ref{fig:str_avg_ro}). The reason for the behavior can be explained from the overhead histogram in Figure \ref{fig:str_ro_hist}. The distance-based selects a high number of high latency hosts for data placement. However, it can find a fairly good number of low-latency hosts at some point in the given environmental setting. Depending on the number of requests the Matchmaker receives, the decision-making time will vary due to contention. This results in the lowering of average replica overhead. Latency-based and spatial-based select low latency hosts initially, leading to fewer replica calls. Once the existing replicas are exhausted, both move to high-latency hosts. This selection from low latency to high latency hosts resulted in getting the observed replica overhead. The dip in histogram around 40-60ms for latency-based and spatial-based shows that a few replicas were selected with less contention at the Matchmaker. As the number of hosts, producer and consumers increase, we believe the overhead difference will become more evident as shown in section \ref{ssec:dec_time}.

        \subsubsection{Average consumer replica selection time}
            \label{ssec:dec_time}
\begin{figure}
                \centering
                \includegraphics[scale=0.4]{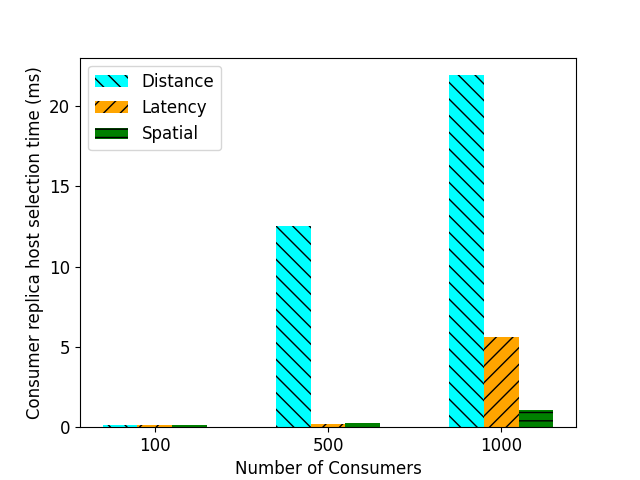}
                \caption{Average consumer replica selection time overhead. The number of hosts and producers is set to 5000 and 50, respectively. The replica selection overhead increases at the Matchmaker with the number of consumers concurrently requesting replicas. Spatial-based can focus on the pruned search space resulting in less time.}
                \label{fig:str_cons_host}
            \end{figure}

            This simulation experiment considers three configurations with 5000 hosts, 50 producers, and varying consumers (100-1000). The experiment aims to observe the average replica selection time per consumer at the Matchmaker. It can be seen in Figure. \ref{fig:str_cons_host} that the time taken to identify a replica for a consumer is the lowest for spatial-based strategy compared to latency-based and distance-based. The reason for this reduction is the pruning of search space for identifying potential host candidates in the Spatial-based. Latency-based makes a near-optimal selection of host in the given setting leading to less replication and hence less contention to access the shared information, resulting in a low selection time compared to distance-based.

    Based on the simulation results, the distance-based strategy incurs more end-to-end latency than latency-based and spatial-based. The same trend is observed for the average number of replicas per producer. For a sparse edge environment, using latency-based or spatial-based can be beneficial. For a dense edge environment, spatial-based strategies' search space pruning can reduce the host selection time per consumer resulting in less replica overhead and hence scale with consumer demand.

        \section{Related Work}
    Over the past five years, multiple approaches have been proposed for data placement at the edge. iFogStor \cite{bib:ifogstor} models the data placement as GAP \cite{bib:gap}, an NP-Hard problem. Its main goal is to identify a single replica storage node where data from the producer be kept to minimize the overall latency between producers and consumers. As the solution cannot scale, a heuristic based on geographical zoning is proposed. The single replication of data may only sometimes be suitable as there can be increased requests resulting in network and storage throttling. Also, if there is an inter-regional flow of data, the zone-specific solution may be sub-optimal. To solve this problem, iFogStorG \cite{bib:ifogstorg} proposed a divide-and-conquer approach. It divided the entire edge infrastructure into several separate and balanced parts to ensure minimized data flow across parts. Within each identified part, the iFogStor approach was run to get the local decision and then combined to get suitable global placement. Here, only a single replica is associated with a producer making it unsuitable for high-request scenarios. To resolve the single replica issue, iFogStorM \cite{bib:ifogstorm} was proposed. The model adds a constraint allowing more than one replica per producer. Similar to iFogStor, iFogStorM cannot be solved in polynomial time. Hence the authors proposed the MultiCopyStorage heuristic, which greedily allows a consumer to select the low latency node among the replicas. It also curbs the replica count when increasing the count does not significantly impact the overall latency. One issue with this technique is that replicas may be too high in an already restricted edge environment. Also, focusing on replicas too far away from the consumer may not be required as edge application users are mostly colocated (autonomous vehicles, AR/VR games). \cite{bib:ifogstors} introduces iFogStorS for small infrastructures that use the shortest path between producers and consumers; and iFogStorP for large infrastructures that use P-median \cite{bib:pmedian} to place P replicas. Compared to MultiCopyStorage, which in parallel sends updates from producer to replicas, iFogStorS/P sends data to one replica, which in turn updates others.
    
    Scientific workflows usually have massive generated datasets stored across multiple cloud data centers leading to high transmission delay. \cite{bib:timedriven} proposes a genetic, self-adaptive, discrete particle swarm optimization data placement strategy (GA-DPSO) that utilizes both the cloud and the edge. The approach does not consider the highly heterogeneous feature of edge nodes. \cite{bib:dedpso} considers the storage capacity at each edge site and the data transmission cost across cloud nodes to make a placement decision. They propose a discrete particle swarm optimization with differential evolution to identify the locations where shared data across multiple scientific workflows can be placed to minimize the transmission time.

    In \cite{bib:coin}, the switches and data indices of unstructured data are associated with coordinates in a virtual space. The data index is stored on servers connected to a switch closest to the virtual space. In \cite{bib:datapop}, inspired by \cite{bib:coin}, the data is placed at the center of a dense network in a virtual space to ensure a shorter distance to all the areas in a region followed by popularity-based replica placement. \cite{bib:joint} jointly place tasks and data, where each block of data is assigned a popularity value to help decide on data placement.

    FogStore \cite{bib:fogstore}, a key-value store, places a set of replicas within the vicinity of the clients and another set of replicas away from the clients to ensure fault tolerance. In addition, it provides differential consistency for data depending on the situation awareness of applications. DataFog \cite{bib:datafog} is a data management platform for IoT which uses spatial proximity to identify the location of replicas. Like FogStore, it also keeps a few replicas in remote locations for fault tolerance. EdgeKV \cite{bib:edgekv} is a decentralized storage system for general-purpose tasks with fault tolerance, reliability guarantees, and strong consistency. Distributed Hash tables are used in EdgeKV to identify locations to store data across different edge nodes.

    Data placement is a well-researched topic in distributed systems \cite{bib:globalplace, bib:p2pscale, bib:stork, bib:greedycover, bib:dynrep, bib:overlay, bib:rcfile, bib:distreplica, bib:saga,bib:unevenreplica}. They mainly focus on reducing the network latency, proximity of dedicated servers to clients, data popularity, adapting to dynamic workloads, partitioning data to adapt to server sizes, and dynamically configuring replicas based on application requirements. Many of the existing distributed databases \cite{bib:dynamodb, bib:cassandra,bib:voldemort} use consistent hashing to store data across different nodes in a load-balanced manner.
        \section{Conclusion}
    \label{sec:conclude}
    Data placement at the edge is a significant challenge that should be addressed to meet the demands of edge applications. The selection of the host node by utilizing location, latency, and spatial awareness can lead to less decision-making time and reduce the end-to-end latency observed by the end user. We compared three data placement strategies: distance-based, latency-based, and spatial-based, using a data placement framework under the same system constraints. The simulation experiments showed that the spatial-based strategy could achieve low end-to-end latency, average replica count, and decision-making time compared to the distance-based strategy. Furthermore, spatial-based is on par with latency-based in terms of end-to-end latency and the average number of replicas in most cases. We also saw that the spatial-based strategy could identify new replica locations with low overhead as the number of consumers increases in a dense edge environment, meaning it can scale with consumer demand.

    The placement of data in the presence of producer mobility, fairness of replica creation based on application requirements, and inclusion of application-specific latency threshold in problem formulation are venues we will be exploring in the future. We also plan to investigate the different types of workloads, and edge server traces for compute, storage, and network to ensure the data placement strategies are adaptable across different environmental settings.

        \bibliographystyle{IEEEtran}
        \bibliography{IEEEabrv, tex/ref.bib}

\end{document}